\documentclass[11pt]{article}
\usepackage{amsmath,amssymb,amsfonts,amstext}
\usepackage{tabularx}
\usepackage{hyperref}
\usepackage{epsf}
\usepackage{graphicx}
\usepackage{amsfonts}
\usepackage{subfigure}
\usepackage[usenames]{color}

\newcommand{\ncd}{\newcommand}
\ncd{\mrm}    {\mathrm}
\ncd{\beq} {\begin{equation}}
\ncd{\eeq} {\end{equation}}
\ncd{\nn}{\nonumber}

\ncd{\rred}{\color[rgb]{1.0,0.1,0.5}}      
\ncd{\ggreen}{\color[rgb]{1.0,0.1,0.5}}
\ncd{\oorange}{\color[rgb]{1,.65,.25}}
\ncd{\bblue}{\color[rgb]{1,1,.1}}
\ncd{\ggrey}{\color[rgb]{0.8,0.8,0.8}}
\ncd{\wwhite}{\color[rgb]{1,1,1}}

\definecolor{Blue}{rgb}{0,0.08,0.45}
\definecolor{Magenta}{cmyk}{0.1,0.8,0,0.1}
\definecolor{Orange}{rgb}{1,0.5,0}
\begin{document}

\title{On the Lyth bound and single field slow-roll inflation} 
%\bigskip
\author{ Gabriel Germ\'an\footnote{gabriel@fis.unam.mx}\\
%EndAName
\\
{\normalsize \textit{Instituto de Ciencias F\'isicas,} }%\\
{\normalsize \textit{Universidad Nacional Aut\'onoma de M\'exico,}}\\
{\normalsize \textit{Apdo. Postal 48-3, 62251 Cuernavaca, Morelos, M\'{e}xico.}}\\
\\
}
\date{}
\maketitle

\begin{abstract}
We take a pragmatic, model independent approach to single field slow-roll inflation by imposing conditions  to the 
slow-roll parameter $\epsilon$ and its derivative $\epsilon^{\prime }.$ To accommodate the recent (large) values of $r$ 
reported by the BICEP2 collaboration we advocate for a decreasing $\epsilon$ during most part of inflation. However 
because at $\phi_{\mathrm{H}}$, at which the perturbations are produced, some $50$ $-$ $60$ e-folds before the end of 
inflation, $\epsilon$ is increasing we thus require that $\epsilon$ develops a maximum for $\phi > \phi_{\mathrm{H}}$ and 
then decrease to small values where most e-folds are produced. The end of inflation might occur trough a hybrid field and a 
small $\Delta\phi$ is obtained with a sufficiently thin $\epsilon$ which, however, should not conflict with the curvature 
of the potential measured by the second slow-roll parameter $\eta$. The conclusion is that under these circumstances 
$\Delta\phi$ and the spectral index $n_{\mathrm{s}}$ are restricted to narrow windows of values.
\end{abstract}

\bigskip
\bigskip

The recent discovery of a large tensor by the BICEP2 \cite{BICEP2} experiment has thrown a lot of excitement in the 
community working in inflationary models mainly because the discovery of the tensor $r$ is taken as a very distinctive 
feature of inflation. Usually models of inflation are studied by considering a potential motivated, in the 
best cases, by an underlying particle theory. The value of $r$ reported by BICEP2 takes the scale of inflation 
to energies similar to the $GUT$ scale where, however we do not have a well understood model of particle physics. Instead 
of pretending the study of a whole potential one can try to study just a small part of it using the information available. 
With this in mind we here study characteristics not of the potential but of the slow-roll parameter $\epsilon$ and its 
derivative $\epsilon^{\prime }$ together with the generation of sufficient e-folds and the Lyth bound. We find that the 
window for single field inflation to occur within the slow-roll approximation narrows. We also find that the spectral index 
$n_{\mathrm{s}}$ is restricted to a narrow window of values during observable inflation.

Let us denote by the subscript ${\mathrm{H}}$ values of quantities at $\phi_{\mathrm{H}}$, at which the perturbations
are produced, some $50$ $-$ $60$ e-folds before the end of inflation. In view of BICEP2 results, the tensor index 
$r_{\mathrm{H}}$ is large: $r_{\mathrm{H}}=0.2^{+0.07}_{-0.05}$ ($r_{\mathrm{H}}=0.16^{+0.06}_{-0.05}$ when foreground 
subtraction based on dust models has been carried out \cite{BICEP2}). For definiteness we take in what follows 
$r_{\mathrm{H}}=0.16$. In the slow-roll single field approximation, $r=16\epsilon$ thus correspondingly we take 
$\epsilon_{\mathrm{H}}=0.01$. The number of e-fold involves integrating over the function $1/\sqrt{2\epsilon}$, thus a 
large $\epsilon$ produce few e-folds around $\epsilon_{\mathrm{H}}.$ 

For an increasing $\epsilon$ during the first few e-folds of observable inflation the Lyth bound \cite{Lyth:1997} implies 
a relatively large $\Delta\phi$. For $N\approx 4$ and $\epsilon\ge r_{\mathrm{H}}/16$, 
$\Delta_4\phi \approx 4\sqrt{2\epsilon} \geq 4 \sqrt{r_{\mathrm{H}}/8} M \approx 0.56 M$. In the Boubekeur-Lyth bound 
\cite{Boubekeur:2005zm} a stronger result follows when $\epsilon$ does not decrease during inflation. Thus it seems that 
we are invited to consider a decreasing $\epsilon$ during most part of inflation with a large number of e-folds generated 
not around $\phi_{\mathrm{H}}$ but close to the end of inflation at $\phi_e$ (for related work see e.g., 
\cite{BenDayan:2009kv}, \cite{Hotchkiss:2011gz}, \cite{Antusch:2014cpa}). This suggest that inflation is terminated 
not by the inflaton-field itself but by some other mechanism e.g, a hybrid field. Let us concentrate in the inflationary 
period without pretending to determine the whole inflationary potential, after all inflation seems to be a transient 
phenomenon in the evolution of the early universe. 

\begin{figure}[!ht]
\begin{centering}
\includegraphics[scale=0.75]{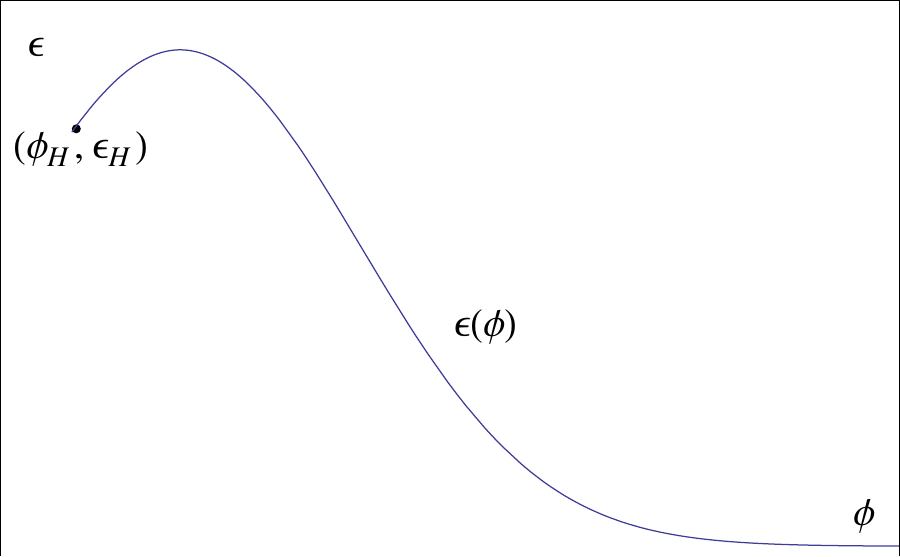}
\caption{\small Plot of $\epsilon(\phi)$ during the inflationary era for a slow-roll parameter $\epsilon$ with a maximum 
at $\phi_{max}$, close to $\phi_{\mathrm{H}} $, (see Figure~\ref{f2}). The maximum is required because 
$\epsilon_{\mathrm{H}}$ is increasing at $\phi_{\mathrm{H}}<\phi_{max}$ and we propose a decreasing $\epsilon(\phi)$ 
(for $\phi>\phi_{max}$) where practically all of inflation occurs. The contribution to the number of e-folds when 
$\epsilon$ can not be bigger tha 5 e-folds for $\Delta\phi$ less than one (see the right hand side of 
Eq.~(\ref{Deltabounded3})). The end of inflation for vanishing $\epsilon$ is triggered by a hybrid field and a small 
$\Delta\phi$ is obtained when $\epsilon$ is sufficiently thin which, however, should not conflict with the curvature of 
the potential measured by the other slow-roll parameter $\eta$.} 
\label{f1}
\end{centering}
\end{figure}

Let us consider the usual slow-roll parameters \cite{Liddle:2000cg} which involve the potential and its derivatives  
\begin{equation}
\epsilon \equiv \frac{M^{2}}{2}\left( \frac{V^{\prime }}{V }\right) ^{2},\quad
\eta \equiv M^{2}\frac{V^{\prime \prime }}{V},
\label{Slowpara}
\end{equation}%
where primes denote derivatives with respect to $\phi$. $M$ is the reduced Planck mass $M=2.44\times 10^{18} \,
\mathrm{GeV}$, we set $M=1$ in what follows. In the slow-roll approximation the observables are given in terms of the 
usual slow-roll parameters \cite{Liddle:2000cg} as follows 
\begin{eqnarray}
n_{\mathrm{t}} &=&-2\epsilon =-\frac{r}{8} , \label{Slowr} \\
n_{\mathrm{s}} &=&1+2\eta -6\epsilon ,  \label{Slowns}  
\end{eqnarray}
where $n_{\mathrm{t}}$ is the tensor spectral index, $r$ is the usual tensor index or the ratio of tensor to scalar 
perturbations and $n_{\mathrm{s}}$ the scalar spectral index. From Eq.~(\ref{Slowpara}) the derivative of $\epsilon$ is 
given by
\begin{equation}
\epsilon^{\prime } = \frac{V^{\prime }}{V}\left( \eta -2\epsilon\right).
\label{epsprima1}
\end{equation}%
Let us concentrate for the moment in the expression where $ V^{'}<0$ with the potential $V$ a monotonically decreasing 
function of $\phi$ during inflation. In this case $\phi$ is evolving away from the origin, thus the derivative of 
$\epsilon$ is
\begin{equation}
\epsilon^{\prime } = -\sqrt{2\epsilon}\left( \eta -2\epsilon\right).
\label{epsprima2}
\end{equation}%
The case $ V^{'}>0$ would correspond to $\phi$ evolving towards the origin and can be analyzed in a similar way. For 
illustrative purposes let us consider $r_{\mathrm{H}}=0.16$, thus $\epsilon_{\mathrm{H}}=0.01$ and from Eq.~(\ref{Slowns}) 
$\eta_{\mathrm{H}}=0.01$ where we have used $n_{\mathrm{sH}}=0.96$ \cite{Ade:2013uln}. At $\phi_{\mathrm{H}}$ we see that
\begin{equation}
\epsilon^{\prime }_{\mathrm{H}} > 0.
\label{epspos}
\end{equation}%
Even for the lowest value reported by BICEP2 ($r_{\mathrm{H}}=0.11$) we get $\epsilon^{\prime }_{\mathrm{H}} > 0$. 
Thus at $\phi_{\mathrm{H}}$ $\epsilon$ is an increasing function of $\phi$. However we want to have a {\it decreasing} 
$\epsilon$. This would suggest that $\epsilon$ has to go through a maximum at $\phi_{max} > \phi _{\mathrm{H}}$ before 
starting to decrease up to a point where $1/\sqrt{2\epsilon}$ has generated sufficient e-folds and inflation is terminated 
by a waterfall field. Thus $\epsilon$ would look like in Figure~\ref{f1}.
\begin{figure}[!ht]
\begin{centering}
\includegraphics[scale=0.75]{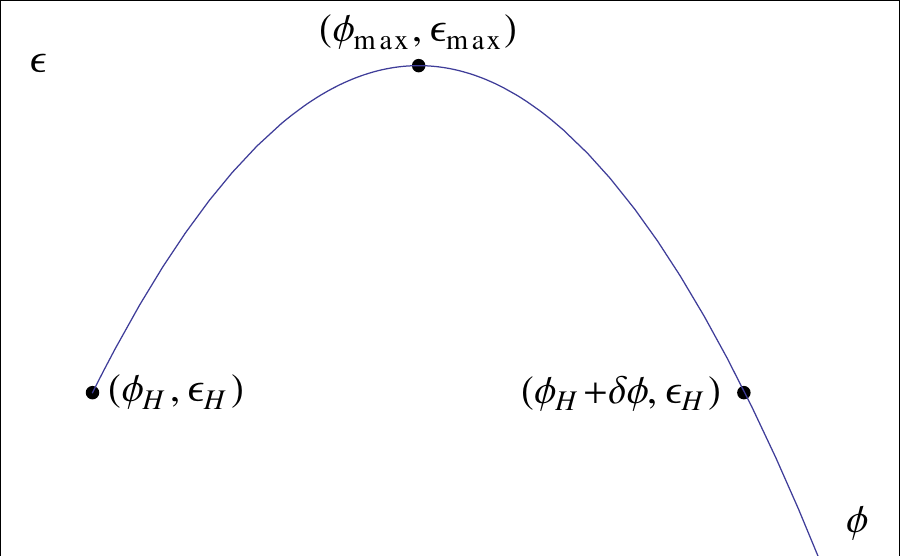}
\caption{\small Plot of $\epsilon$ as a function of $\phi$, this is a zoom of Figure~\ref{f1} around the maximum of 
$\epsilon(\phi)$. Typically $\epsilon_{\mathrm{H}}$ is close to $\epsilon_{max}$ because  
(see Eq.~(\ref{epsprima2})) $\epsilon^{\prime }_{\mathrm{H}}\approx 0.0014$. Thus, although $\phi_{\mathrm{H}}$ is 
located left of $\phi_{max}$ the contribution to the number of e-folds during the evolution $\delta\phi/2$ where 
$\epsilon$ is increasing is small.} 
\label{f2}
\end{centering}
\end{figure}
The value $\phi _{\mathrm{H}}$ at which $n_{\mathrm{sH}} =0.96$ lies to the left of $\phi_{max}$ so that 
$\epsilon^{\prime }_{\mathrm{H}}$ is positive although small. We observe that an $\epsilon$  like shown in 
Figure~\ref{f1} has the potential to generate large values of $r_{\mathrm{H}}$ while sufficient inflation is produced away 
from $\phi_{\mathrm{H}}$. Close to  the end of inflation at $\phi_{e}$ the potential becomes very flat and a hybrid 
mechanism should terminate inflation. A small $\Delta\phi$ is obtained with a sufficiently thin $\epsilon$, thus 
$\epsilon$ should not only be tall with small end but also thin. We would expect that a thin $\epsilon$ conflict with the 
slow-roll parameter $\eta$ which measures the curvature of the potential. 
After $\phi$ has reached the maximum $\epsilon^{\prime}$ becomes negative and we can rewrite Eq.~(\ref{epsprima2}) in a 
more convenient form as
\begin{equation}
\eta=2\epsilon-\frac{\epsilon^{\prime } }{\sqrt{2\epsilon}}, \quad \quad \epsilon^{\prime } < 0, 
\quad \quad \phi > \phi_{max}.
\label{eta}
\end{equation}%
Note that both terms $2\epsilon$ and $-\frac{\epsilon^{\prime } }{\sqrt{2\epsilon}}$ in Eq.~(\ref{eta}) are positive thus 
$0<\eta<1$ during inflation. The first term is negligible w.r.t. the second because we want a large $\epsilon^{\prime}$ 
and $\epsilon \leq 0.01$, thus for $\phi > \phi_{max}$,  $2\epsilon$ decreases while $-\frac{\epsilon^{\prime } }{\sqrt{2\epsilon}}$ grows large
\begin{equation}
\eta=2\epsilon-\frac{\epsilon^{\prime } }{\sqrt{2\epsilon}} \approx -\frac{\epsilon^{\prime } }{\sqrt{2\epsilon}} 
\approx \frac{-\Delta\epsilon}{\Delta\phi\sqrt{2\epsilon}}.
\label{etaapprox}
\end{equation}%
Thus
\begin{equation}
\Delta\phi \approx \frac{-\Delta\epsilon}{\sqrt{2\epsilon}}\, \frac{1}{\eta} > \frac{-\Delta\epsilon}{\sqrt{2\epsilon}} > 
\frac{\epsilon_{\mathrm{H}}-\epsilon_{\mathrm{N}}}{\sqrt{2\epsilon_{\mathrm{H}}}} 
\approx \left(\frac{\epsilon_{\mathrm{H}}}{2}\right)^{\frac{1}{2}}.
\label{Deltafi}
\end{equation}%
For $\epsilon_{\mathrm{H}}=0.01$ (corresponding to $r_{\mathrm{H}}=0.16$ when $n_{\mathrm{sH}}=0.96$) we get
\begin{equation}
\Delta\phi  > 0.07,
\label{Deltafi07}
\end{equation}%
this is a lower bound for $\Delta\phi$ just compatible with $\eta<1$. Here $\phi>\phi_{max}$ and $\epsilon$ is decreasing. 

If we were not considering any other observational input but only the fact that at $\phi_{\mathrm{H}}$, $n_{\mathrm{sH}}=0.96$ 
and $r_{\mathrm{H}}=0.16$ (thus $\epsilon_{\mathrm{H}}=0.01$ and $\eta_{\mathrm{H}}=0.01$), then in a model with a thin 
$\epsilon$ the contribution to the number of e-folds from $\phi_{\mathrm{H}}$ to $\phi_{\mathrm{H}}+\delta\phi$ 
(see Figure~\ref{f2}) would be negligible and $N$ would be due to e-folds generated with a {\it decreasing} 
$\epsilon(\phi>\phi_{max})$ where $\epsilon^{\prime} <0$. During most of this era $\epsilon < r_{\mathrm{H}}/16$ and the 
Lyth bound would be modified as
\begin{equation}
\Delta_4\phi \approx 4\sqrt{2\epsilon}<4\sqrt{\frac{r_{\mathrm{H}}}{8}}\approx 0.56 .
\label{Deltafi4}
\end{equation}%
Thus the inequality is inverted, together with Eq.~(\ref{Deltafi07}) we would get
\begin{equation}
0.07 < \Delta\phi < 0.56 .
\label{Deltabounded1}
\end{equation}%
There is, however evidence that the power-spectrum over this range of scales has been observed to be decreasing in 
amplitude as the scales decrease, which means that, while this range of scales were leaving the horizon during inflation, 
$\epsilon$ was increasing\footnote{We thank Shaun Hotchkiss for correspondence on this point.} and the Lyth bound seems to be an 
inevitable consequence. If we incorporate this observation as an input we see that the inequality Eq.~(\ref{Deltabounded1}) 
should be modified as follows. Defining the quantity $\delta _{\mathrm{ns}}$ by 
$\delta_{\mathrm{ns}}\equiv 1-n_{s}$, we note that Eq.~(\ref{epsprima2}) can be written as
\begin{equation}
\epsilon^{\prime } = \frac{1}{2}\sqrt{2\epsilon}\left( \delta _{\mathrm{ns}} -2\epsilon\right).
\label{epsprima3}
\end{equation}%
Requiring that $\epsilon$ increases during observable inflation (i.e., from  $\phi_{\mathrm{H}}$ to $\phi_{max}$) implies 
$\epsilon^{\prime }>0$ during this range, or $\delta _{\mathrm{ns}} > 2\epsilon > 0$, that is $\delta _{\mathrm{ns}} >0$ 
which means that during observable inflation the spectral index should be less than one, $n_{\mathrm{s}}  < 1$, also 
$\frac{1}{2\epsilon}   > \frac{1}{\delta _{\mathrm{ns}} } $. From here follows that 
$dN = \frac{d\phi}{\sqrt{2\epsilon}}   > \frac{d\phi}{\sqrt{\delta _{\mathrm{ns}} }},$ or
\begin{equation}
\Delta\phi \leq \Delta N \sqrt{\delta _{\mathrm{ns}} } \,.
\label{enese}
\end{equation}%
Combining with the Lyth bound
\begin{equation}
\Delta N\sqrt{2\epsilon_{\mathrm{H}}} \leq \Delta\phi \leq \Delta N \sqrt{\delta _{\mathrm{ns}}}\,.
\label{Deltabounded3}
\end{equation}%
It follows that $\delta _{\mathrm{ns}}\geq 2\epsilon_{\mathrm{H}} \approx 0.02$, thus $n_{\mathrm{s}}  < 0.98$ or 
\begin{equation}
n_{\mathrm{sH}}  < n_{\mathrm{s}}  < 0.98,
\label{nsbounded}
\end{equation}%
during observable inflation. If we take $n_{\mathrm{sH}} =0.96$ then Eq.~(\ref{Deltabounded3}) becomes
\begin{equation}
0.56 \leq \Delta_4\phi \leq 0.8.
\label{Deltabounded4}
\end{equation}%
Thus allowing for and increasing $\epsilon$ during observable inflation with a posterior decrease we should be able to 
construct models where $\Delta\phi\leq 1$. We note from the right hand side of Eq.~(\ref{Deltabounded3}), 
that keeping $\Delta\phi\leq 1$ can only be done for $\Delta N \leq \frac{1}{\sqrt{\delta _{\mathrm{ns}}}}\approx 5$ e-folds.

We have shown that by imposing conditions coming from the large tensor $r$ reported by BICEP2 to the slow-roll parameter 
$\epsilon$ and its derivative $\epsilon^{\prime }$ we can accommodate sufficient inflation for  $\Delta\phi$ no bigger 
than one if $\epsilon(\phi)$ is a function with a maximum, thin and decreasing to vanishing values close to the end of 
inflation at $\phi_e$. The maximum is required because $\epsilon_{\mathrm{H}}$ is increasing at 
$\phi_{\mathrm{H}}$ and then $\epsilon(\phi)$ should decrease for $\phi>\phi_{max}$ where practically all of 
inflation occurs. The contribution to the number of e-folds when $\epsilon$ is growing can be at most 5 e-folds for 
$\Delta\phi$ less than one. The end of inflation for vanishing $\epsilon$ can be triggered by a hybrid field and a small 
$\Delta\phi$ is obtained when $\epsilon$ is sufficiently thin which, however, should not conflict with the curvature of 
the potential measured by the other slow-roll parameter $\eta$. Under these circumstances $\Delta\phi$ and the spectral 
index $n_{\mathrm{s}}$ are restricted to narrow windows of values. Our main results are given by 
Eqs.~(\ref{Deltafi07}), ~(\ref{nsbounded}) and ~(\ref{Deltabounded4}).

\section{Acknowledgements}

We gratefully acknowledge support from \textit{Programa de Apoyo a Proyectos de Investigaci\'on e Innovaci\'on 
Tecnol\'ogica} (PAPIIT) UNAM, IN103413-3, \textit{Teor\'ias de Kaluza-Klein, inflaci\'on y perturbaciones gravitacionales}.

\end{document}